\documentclass[prd,11pt,tightenlines,nofootinbib,superscriptaddress]{revtex4}

\usepackage{amssymb,amstext,amsmath}
\usepackage[dvips]{graphicx}
\usepackage{latexsym}
\usepackage{psfrag}
\usepackage{amsfonts}
\usepackage{bbm}
\usepackage{color}

\newcommand{\Ref}[1]{(\ref{#1})}

\newcommand{\eqa}{\begin{eqnarray}}
\newcommand{\neqa}{\end{eqnarray}}
\newcommand{\equ}{\begin{equation}}
\newcommand{\nequ}{\end{equation}}

\def\ra{\rangle}
\newcommand{\ket}[1]{|{#1}\ra}

\def\d{\delta}
\def\f{\frac}

\def\wtl{\widetilde}

\newcommand{\n}{\nabla}

\newcommand{\scr}{\rm\scriptscriptstyle}

\newcommand{\SU}{\mathrm{SU}}

\newcommand{\Spin}{\mathrm{Spin}}

\let\th=\theta

\newcommand{\lp}{\ell_{\rm P}}

\newcommand{\be}{\begin{equation}}
\newcommand{\ee}{\end{equation}}

\begin{document}

\title{\Large\bf Sub-leading asymptotic behaviour of area correlations in the
 Barrett-Crane model}

\author{J. Daniel Christensen}\email{jdc@uwo.ca}
\affiliation{Department of Mathematics, University of Western Ontario, London, ON N6A 5B7, Canada}

\author{Igor Khavkine}\email{i.khavkine@uu.nl}
\affiliation{Institute for Theoretical Physics, Utrecht University, 3584 CE Utrecht, The Netherlands}

\author{Etera R. Livine}\email{etera.livine@ens-lyon.fr}
\affiliation{Laboratoire de Physique, ENS Lyon, CNRS-UMR 5672, 46 All\'ee d'Italie, Lyon 69007, France}

\author{Simone Speziale}\email{simone.speziale@cpt.univ-mrs.fr}
\affiliation{Centre de Physique Th\'eorique, CNRS-UMR 6207, Campus de Luminy, Case 907, Marseille 13009, France}

\begin{abstract}

The Barrett-Crane spin foam model for quantum gravity provides an excellent setting
for testing analytical and numerical tools used to probe spinfoam models.
Here, we complete the report on the numerical
analysis of the single 4-simplex area correlations begun in Phys.\ Lett.\ B670 (2009) 403--406, discussing the
next-to-leading order corrections (``one-loop'' corrections) with particular attention to
their measure dependence, and the difference between the Gaussian and Bessel
ans\"atze for the boundary state.
\end{abstract}

\maketitle


\section{Introduction}\label{se:intro}

The current challenge for spinfoam models of quantum gravity \cite{book} is the understanding of their semiclassical regime: how to recover general relativity and the subsequent quantum corrections due to the quantum fluctuations of the geometry (around the flat space-time) from the spinfoam amplitudes. A key proposal by Rovelli \cite{RovelliProp} for computing the graviton propagator through correlations of geometrical observables has provided a well-suited tool to probe the semi-classical behaviour of spinfoam models \cite{grav2,grav3,Alesci,Alesci2,letter,Bianchi} 
(see also \cite{Speziale} for a review).

The most studied spinfoam model until recently was the Barrett-Crane model \cite{BC,BC2}. The corresponding spinfoam amplitudes are built from the $10j$-symbol first defined in \cite{BC}. Its asymptotic limit, relevant to the semi-classical behaviour of the model, has been well-studied analytically \cite{BaCh,BCHT,asymptBW,asymptBS,asymptBCE,asymptFL}. The leading order is dominated by degenerate geometries \cite{asymptBS,asymptBCE,asymptFL} and the contribution corresponding to non-degenerate configurations is an oscillating sub-leading term of the asymptotic expansion. The key idea of Rovelli's proposal was  to kill the degenerate leading order through destructive interference with the boundary state and then compute the ``graviton propagator" using the proper non-degenerate oscillating term \cite{RovelliProp}. This was rigourously implemented in \cite{grav3} where the graviton propagator was computed analytically for a single 4-simplex at leading order. These results were checked numerically in \cite{letter}, which provided the first numerical confirmation of the sub-leading oscillatory behaviour of the $10j$-symbol. There was a perfect fit with the analytical predictions of \cite{grav3} at leading order in the asymptotic regime. Moreover the numerical simulations allowed us to probe the small scale behaviour of the Barrett-Crane model, showing that the correlations were regularized at the Planck scale. Indeed, although the correlations were found to decrease as expected as the inverse squared distance at large scale (above a hundred Planck lengths), they reached a maximum at a few Planck lengths when going to the zero-distance limit and then decreased again. This correlation peak provides a dynamical definition of a minimal distance and shows that the gravitational correlations are regularized at short scales (at least when looking at a single 4-simplex).

Although the Barrett-Crane model gives this remarkable result, it does not lead to the right semi-classical behaviour. Indeed, although the correlations properly fall as $1/r^2$ at large distances, they do not have the correct tensorial structure \cite{grav2,Alesci}. This opened the door to the new EPR-FK-LS class of spinfoam models which are better-behaved semi-classically and are based on a relaxed imposition of the so-called simplicity constraints of the Barrett-Crane model. These have been constructed algebraically \cite{EPR1, EPR2}, with coherent states techniques \cite{coherent1,FK,coherent2}, and using discrete path integral techniques \cite{FC,valentin}. This new EPR-FK-LS spinfoam vertex (for the quantum 4-simplex) was shown analytically and numerically to propagate coherent wave packets of 3d geometry \cite{EPRwave}. Its asymptotic behaviour improves on that of the Barrett-Crane vertex \cite{asymptEPR}, and it is related to Regge calculus in area-angle variables (see \cite{Dittrich}). Finally, the graviton propagator was computed in this framework for a single 4-simplex and was found to yield the correct tensorial structure at leading order \cite{EPRgraviton}. Therefore, these EPR-FK-LS spinfoam models have a better semi-classical behaviour than the previous Barrett-Crane model and there is much hope that we will recover classical general relativity plus quantum gravity corrections in a continuum limit.

On the other hand, the Barrett-Crane amplitudes still provide a sensible semiclassical dynamics when restricted to area correlations on a single 4-simplex. And being much simpler than the EPR-FK-LS amplitudes, they provide a useful tool for numerical and analytical studies. Their simplicity comes from the absence of intertwiner degrees of freedom, which is ultimately also the reason why we can not look at more generic geometric correlations nor properly glue more 4-simplices together. The model is also the simplest non-topological spinfoam model for 4d quantum gravity, thus an arena for studying issues like semiclassical symmetries and renormalization (under coarse-graining) in the formalism.

In the present paper, we push further the numerical work started in \cite{letter}. In  the next section, we start by reviewing the definition of the Barrett-Crane model and the ambiguity in defining the path integral measure. We also present a plot showing the degenerate contribution and two other sub-leading contributions which also mask the expected oscillations. Then, in Section \ref{se:corr}, we define the area-area correlations. We restrict ourselves to factorized (homogeneous) boundary states. We recall the two types of ansatz for such states: the (phased) Gaussian ansatz originally introduced in \cite{RovelliProp} and the Bessel ansatz later defined in \cite{3d,grav3}. The Bessel state is simply a refinement of the Gaussian ansatz allowing more precise analytical computations but the two states are asymptotically the same. We then present our numerical results, plots of the correlations in term of the area scale $j_0$ for different choices of the Gaussian width $\alpha$ and the parameter $k$ labeling the ambiguity in the path integral measure (more precisely the face amplitude). We recover the typical behaviour already identified in \cite{letter}: a plot with a peak for some given small value of $j_0$ and then the correlations going down as $1/j_0$ for large $j_0$. The precise leading order coefficient in front of $1/j_0$ depends on $\alpha$ but is not sensitive to $k$ while the precise position of the peak depends on both $\alpha$ and $k$. We also find a new oscillatory regime for intermediate values of $j_0$, which is rather unexpected.

In Section \ref{nlosection}, we focus on the next-to-leading order (NLO) correction in the asymptotic regime, making use of new numerical methods~\cite{Khavkine} that allow us to obtain the required precision. We find a sub-leading term that goes like $1/j_0^2$. We estimate the coefficient for different choices of $\alpha$ and $k$ and we conjecture a quadratic dependence on the measure ambiguity $k$. This NLO correction should correspond to a one-loop correction to the graviton propagator and Newton potential (which defines the $\beta$-function for Newton's constant at first order). In Section \ref{sectionpeak}, we discuss the small scale behaviour, interpreting the peak as defining a dynamical notion of a minimal distance. In Section \ref{besselsection}, we turn to the Bessel state which introduces further interferences and enhances the oscillatory regime for small/intermediate values of $j_0$. Finally, in Section \ref{se:concl}, we discuss the relevance of our choice of boundary state and the limits of the power of numerical simulations for the study of spinfoam correlations.
We end with an appendix which gives a summary of the new numerical techniques which make the NLO calculations possible.

\section{Barrett-Crane dynamics}\label{se:BC}

The object that we study is the area correlation on a single Riemannian 4-simplex, defined in \cite{RovelliProp} and further studied in \cite{grav2,grav3,Bianchi}. This is a 10 by 10 matrix with components
\equ\label{W}
W_{ab}(\alpha,j_0)=\f1{\cal N} \sum_{j_l}
\, {\mathbbm h}(j_a) \, {\mathbbm h}(j_b) \, \Psi_q[j_l] \, K[j_l].
\nequ
In the following we will assume that the reader has a certain familiarity with this quantity. For more details and motivation, as well as a description of its connection to the graviton propagator, we refer to the literature cited above. In a few words, the notation used in \Ref{W} is as follows:
the quantities $j_{l}$ are spins, i.e.\ non-negative half-integers, which label $\SU(2)$ representations.
They are indexed by subscripts $a$, $b$ and $l$ which run over the $10$ triangles of a 4-simplex.
The normalization ${\cal N}$ is the same as the displayed sum, without
the field insertions ${\mathbbm h}(j_a)\equiv\big(A_a^2-A_0^2\big)/A_0^2$, which represent the fluctuation of the area
$A_a=\lp^2 \, (2j_a+1)$ of the triangle $a$ around its background value $A_0=\lp^2 \, (2j_0+1)$.
The function $\Psi_q$ is a boundary state, which should be peaked on the geometry of a regular 4-simplex.
We describe our choices of this state in Section~\ref{IIIA}.
The model-dependent kernel $K[j]$ encodes the dynamics.
In this paper, we focus on the Barrett-Crane model \cite{BC},
and summarize the properties of the kernel in Sections~\ref{IIA} and~\ref{IIB}.

Compared to the general definition (see e.g.\ \cite{grav2}), two important restrictions appear in \Ref{W}:
(i) the expression is the contribution of a single 4-simplex, i.e.\ it is the first approximation
in the $\lambda$ expansion of the group field theory; (ii) the area-angle and angle-angle correlations also needed to reconstruct the graviton propagator have been discarded.
Both restrictions are consistent with the choice of dynamics; in fact, the Barrett-Crane model does not give a consistent semiclassical limit on more than one 4-simplex, and furthermore it carries no dynamics for the intertwiners---the quantum degrees of freedom associated to angles  \cite{grav2,Alesci}. Therefore, sensible results from the BC model can be extracted
only if one restricts attention to area correlations on a single 4-simplex.
This in turn means only looking at the ``diagonal components'' $h_{\mu\mu}(x) h_{\nu\nu}(y)$ of the graviton propagator.
For such components, the BC model proves very useful as a non-trivial test for a number of features of the whole approach based upon general boundary formulas such as \Ref{W}.

\subsection{A class of models}\label{IIA}

In this paper we consider the class of models defined by the following amplitude,
\equ\label{K}
K[j_l] = \Big( \prod_l (2{j_l}+1)^k  \Big) \, \{10j\},
\nequ
where $\{10j\}$ denotes the $10j$-symbol (described further in Section~\ref{IIB}) and
the integer $k$ parameterizes the choice of triangle weight in the measure.
The interest in this class of models comes from the fact that it parameterizes some freedom in the choice of the
functional integration over the $B$ field. This is the analogue of the usual ambiguity in defining the integration
measure over 4-geometries ${\cal D}g_{\mu\nu}$. The spin foam formalism suggests using the measure that gives
exactly a topologically invariant partition function before imposing the constraints reducing BF theory to GR, a choice
which amounts to taking $k=2$ in \Ref{K}.
Nevertheless, different choices have been considered, and it is interesting to study how they affect the physics.
For instance, different values of $k$ change the divergent properties
of the model, as investigated in \cite{Perini} for the new EPR-based models.
Here we study how they affect the next-to-leading order behaviour.

The form \Ref{K} is not the most general BC-like model: one can also consider weights associated to each tetrahedron, in the form of powers of the dimension of the intertwiner space associated to each tetrahedron \cite{daniele}. These weights affect the convergence of the model \cite{BCHT, PerezRov}. Such factors introduce further coupling between the $j_l$'s and complicate both the analytical and numerical studies of the kernel, so we do not consider them
here. Notice however that, as shown in \cite{3d,Io} in the 3d case, both ambiguities should not affect the leading behaviour of the correlations $W$.

\bigskip

\subsection{The $10j$-symbol and its asymptotics}\label{IIB}

The $10j$-symbol $\{10j\}$
is a $\Spin(4)$ invariant tensor constructed with Clebsch-Gordan coefficients. It is the evaluation
of a $\Spin(4)$ spin network and can be expressed as an integral over five copies of $\SU(2)$:
\equ\label{10j}
\{10j\}\,=\, (-1)^{\sum_l 2j_l}\int_{\SU(2)^{5}} \prod_v dg_v\, \prod_l 
\chi_{j_l}(g_{v(l)}^{-1} \, g_{v'(l)}).
\nequ
(For more details see, e.g., \cite{grav3}.)
In this expression we associate one $\SU(2)$ group element $g$ to each
tetrahedron $v$ of the 4-simplex, and $v(l),v'(l)$ are the two tetrahedra sharing the triangle
$l$. Finally
\[
\chi_j(g)=\f{\sin((2j+1)\phi)}{\sin\phi}
\]
is the character of the group element $g$
in the $j$-representation expressed in terms of its class angle $\phi\in[0,2\pi]$.
Notice the phase factor $(-1)^{\sum_l 2j_l}$ in \Ref{10j}, which has been so far discarded in the
spin foam graviton literature~\cite{RovelliProp,grav2,grav3,Alesci,Alesci2,Bianchi}.
We will see in Section~\ref{IIIA} that it simply corresponds to a small change in the choice
of boundary state.

We now briefly recall the known results on the asymptotics of the $10j$-symbol,
referring to the literature for details.
The only type of large spin behaviour that has been studied in detail is
the asymptotics as a fixed set of spins is scaled.  More precisely, let
$j_l$ be ten chosen spins and consider the value of the $10j$-symbol as
the spins are adjusted so that the dimensions $2j_l + 1$ become $\lambda (2j_l+1)$
for $\lambda$ a positive integer.\footnote{As was already shown in the three-dimensional case with the $6j$-symbol \cite{3d2,maite1}, the structure of the asymptotic expansion is simpler when expressed in terms of $2j+1$ rather than $j$, and it was also observed in \cite{asymptBCE} that this leads to better behaviour.}
In~\cite{asymptBW} it was shown that certain non-degenerate configurations
contribute an oscillating Regge term proportional to
\equ\label{10jasymp}
\{10j\} \sim \lambda^{-9/2} (-1)^{\sum_l 2j_l} \cos \Big(\sum_l \lambda (2j_l+1) \th_l + \kappa \frac{\pi}{4} \Big),
\nequ
where $\kappa$ is an integer and the $\th_l$ are the exterior dihedral angles of a flat 4-simplex with areas $A_l=2j_l+1$.  We have suppressed a sum over such 4-simplices if there is more than one possibility.

\begin{figure}[ht]
\includegraphics[width=11cm]{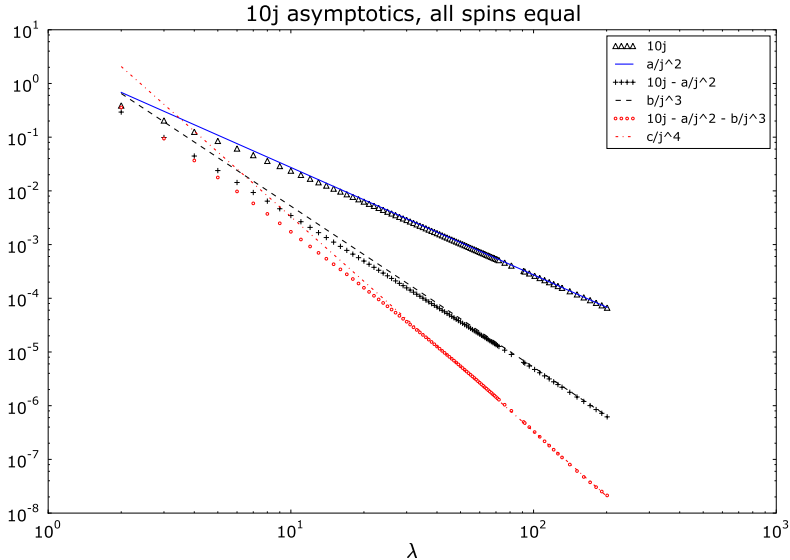}
\caption{\small{Log-log plot of the values of the equilateral $10j$-symbol. 
The dots are the numerical evaluations, the lines the asymptotic fits.
The oscillating behaviour depending upon the Regge action (predicted to scale as $1/\lambda^{9/2}$) is completely masked by the degenerate configurations, whose leading order is approximately $2.73/\lambda^{2}$. After subtraction of this leading order, further non-oscillating contributions appear, which are approximately $-5.20/\lambda^3$ and $33.1/\lambda^4$.}}\label{Fig10j}
\end{figure}

The geometric saddle point \Ref{10jasymp} of the $10j$-symbol is notoriously masked by a non-oscillating term, which comes from degenerate saddle points in \Ref{10j} with no direct geometrical meaning. This term was found to scale like $1/\lambda^2$ and thus dominates the large spin limit of the $10j$-symbol~\cite{asymptBS,asymptBCE,asymptFL}.
The situation is illustrated in Figure \ref{Fig10j}, where the values of the regular $10j$-symbol (with $2 j_l + 1 = \lambda$) are plotted as a function of $\lambda$.

It is interesting to note that even after removing the dominating contribution of order $1/\lambda^2$, there are additional non-oscillating contributions still hiding the Regge term. To see this, in Fig.~\ref{Fig10j} we plot the next-to-leading and next-to-next-to-leading orders, obtained by subtracting a numerical fit of the previous order.
These additional contributions may correspond to those predicted in~\cite{asymptBS}, or they may be the power series corrections of the degenerate saddle point giving the leading order in $1/\lambda^2$.
At the present time, our calculations are not accurate enough to be sure that there aren't additional contributions involving fractional powers of $\lambda$ (as predicted by~\cite{asymptBS}), or to
be able to discern the expected oscillating contribution of order $1/\lambda^{9/2}$.
We hope to address this in the near future.

\bigskip

This situation in which degenerate configurations hide the Regge term prevents a direct numerical confirmation of the asymptotics \Ref{10jasymp}, and it has often been seen as a problem with the Barrett-Crane model. The development of the general boundary formalism and the definition of the graviton propagator as in \Ref{W} has finally allowed us to circumvent this problem, and the numerical results we present below are an important test of \Ref{10jasymp} and in particular of its non-trivial dependence on the Regge action.
These results are relevant for all spin foam models having degenerate configurations emerging in the semiclassical limit, even if those contributions hide the geometric configurations, as is also the case for the new EPR-FK-LS spin foam models. Following the example of the Barrett-Crane model presented here, the degenerate configurations might not be relevant in the study of semiclassical physics with the general boundary formalism.

\section{Numerical study of area correlations}\label{se:corr}
\label{numericalsection}

An advantage of the BC vertex amplitude is that its integral representation \Ref{10j}
can be written as an integral over ten angles \cite{asymptBS, asymptFL},
\equ\label{10angles}
\{10j\}\,=\,(-1)^{\sum_l 2j_l}
\int \prod_l d\phi_l\, C(\phi_l) \, \prod_l \chi_{j_l}(\phi_l).
\nequ
Notice that this expression is almost factorized: the ten angles couple only through the
(non-linear) measure term $C(\phi_l)\equiv \d (\det G[\phi_l] )$, where $G[\phi_l]$ is the Gram matrix of the 4-simplex (see, e.g., \cite{grav3} for more details).
This will play an important role in the following.

\subsection{On the boundary state}\label{IIIA}
At this point we decide to work with the simplest ansatz for the boundary state, namely a factorized boundary state
\equ \label{psiFact}
\Psi_q[j_l] = \prod_l \psi[j_l],
\nequ
where $\psi[j]$ would be peaked around some background value $q=(A_0(j_0), \th)$, where $\th$ is the 
dihedral angle. One could also take in \Ref{psiFact} different states for each representation label $j_l$, but here we make the choice of a homogeneous ansatz with all the $\psi_l$'s equal, which corresponds to the quantization of an equilateral 4-simplex.
The factorization allows a simpler analytical and numerical study of the correlation.
Indeed, using this particular choice of state and the integral expression \Ref{10angles} of the kernel, we can rewrite the area correlations \Ref{W} as
\equ\label{W1}
W_{ab}(k,\alpha,j_0)\,=\, \f{1}{{\cal N}}\int \prod_l d\phi_l\, C[\phi_l]\,
\wtl{w}(\phi_a)\wtl{w}(\phi_b)\prod_{l\ne a,b} w(\phi_l),
\nequ
where the normalization ${\cal N}$ is $\int \prod_l d\phi_l\, C[\phi_l]\,\prod_{l} w(\phi_l)$
and the integrands are
\equ\label{w}
w(\phi)=\sum_j (2j+1)^{k} \, \psi[j] \, \chi_{j_l}(\phi_l), \qquad
\wtl{w}(\phi)=\sum_j (2j+1)^{k} \, {\mathbbm h}(j) \, \psi[j] \, \chi_{j_l}(\phi_l).
\nequ
We see that the factorization assumption \Ref{psiFact} allows us to perform each sum over the spins separately. This turns out to be crucial for improving numerical stability and allowing us to get accurate results.

Concerning the boundary state \Ref{psiFact}, we consider in this paper the two choices
that have so far appeared in the literature:
\begin{itemize}
\item the Gaussian state, introduced in \cite{RovelliProp} and adapted to the factorized form in \cite{letter},
where each factor in \Ref{psiFact} is given by
\equ\label{psiG}
\psi[j] = e^{-\f{\alpha}{2j_0}(j-j_0)^2} \, e^{i\theta(2j+1)}.
\nequ
\item the Bessel-based state introduced in \cite{3d,grav3}, where each factor in \Ref{psiFact} is given by
\eqa\label{psiB}
\psi_{\scr B}(j) = \f{I_{|j-j_0|}(\f{j_0}{\alpha})-I_{j+j_0+1}(\f{j_0}{\alpha})}{ \sqrt{ I_{0}(\f{2j_0}{\alpha}) - I_{2j_0+1}(\f{2j_0}{\alpha})}} \,\cos((2j+1) \theta).
\neqa
Here $|j-j_0|\in{\mathbbm N}$ and $I_{n}(z)$ is a modified Bessel function of the first kind.
\end{itemize}
As we review below in Section \ref{besselsection}, the Bessel part of \Ref{psiB} reduces in the large spin limit
precisely to the Gaussian in \Ref{psiG}. Therefore, at leading order, 
the only difference between \Ref{psiG} and \Ref{psiB} is in the phase, which is complex in the first
case and real in the second case. 
Let us recall that the rationale for choosing in \cite{grav3} the form \Ref{psiB} with its real phase
was to be able to perform the sums \Ref{w} \emph{analytically}, thus turning the definition of the area
correlator into an exact group integral. This allowed us to obtain an integral representation
of the correlator and define the perturbative expansion as a saddle point approximation.
With the original choice \Ref{psiG}, one has to first approximate the sums as integrals, and then perform a saddle point analysis of the integrals. In both cases, it is ultimately the existence of saddle points with a geometric meaning that leads to the desired leading order result.%
\footnote{For these geometric saddle points to exist, it is important that the \emph{interior} dihedral angle $\th = \arccos(1/4)$ of the regular 4-simplex be used  in the Gaussian state~\Ref{psiG}. This is due to the extra phase in \Ref{10j} and \Ref{10jasymp}. Some past work~\cite{RovelliProp,grav2,Alesci,Alesci2,Bianchi} used the exterior angle $\arccos(-1/4)$, but those sources also omitted the phase in front of~\Ref{10j}, and one can show that these two effects cancel out at the leading order of the correlator~\Ref{W} (up to complex conjugation).
On the other hand, the phase $(-1)^{\sum_l 2j_l}$ is irrelevant if the spins all have the same parity. This is the situation if one uses the Bessel-based boundary state \Ref{psiB}, see \cite{grav3}. In this case, both choices of angles in the boundary state give the right asymptotics. More on this below in Section \ref{besselsection}.}

Here we are interested in a numerical evaluation and thus there is no preference for \Ref{psiB},
as the sums \Ref{w} can be performed numerically for both choices.
We first focus on the simpler Gaussian boundary state. We discuss the Bessel state and compare the two choices later in Section \ref{besselsection}.%
\footnote{One advantage of the Bessel choice is to put the integral formulation on more solid grounds, in particular allowing one to prove that the degenerate configurations do not contribute to the perturbative expansion \cite{grav3}. A second one is to make the spin $j_l$ and the dihedral angle $\th_l$ into conjugate variables in the sense of $\SU(2)$ harmonic analysis.}

However, it is important to stress that the assumption of a factorized boundary state is most likely too simple to capture a true physical state.
The example of linearized Regge calculus shows that in general the state will have the same
structure of the kernel \cite{Bianca}, as already suggested in the original paper \cite{RovelliProp}.
However, this factorized boundary state is the only setting in which we are able to perform
numerical simulations (for large length scales). It is thus a technical limitation, and we stress that this is one of the weak points of the present numerical power, which deserves attention for the future.

\subsection{On the numerical methods}
The reason the factorization allows the numerical simulations to be performed much faster is that one can bring the sum over the ten spins in \Ref{W} inside the integral for the kernel $K$, and then decouple them into a product of independent sums.
This decoupling simplifies the numerical task by decreasing the number of sums over $j$
required to compute the correlations. The resulting 9-dimensional integrals are still very difficult to compute numerically, due to their oscillatory nature, but can be estimated using Monte Carlo methods with between $10^9$ and $10^{10}$ sample points.
This method is able to handle moderately high $j_0$, with sufficient accuracy to determine the leading order behaviour. This was used in~\cite{letter} to support the analytic calculations that appeared in the literature, and to give a first glance at the short scale picture.

Here we push the analysis further, and discuss the next-to-leading order and its
dependence on the triangle amplitudes in \Ref{K}.
Unfortunately, the Monte Carlo method converges too slowly to
efficiently extract the next-to-leading order behaviour. An alternative
method must be used to probe these subdominant asymptotics, which are
presented in the next section.  This alternative method 
uses the spin summation formula for the
$10j$-symbol~\cite{CEalg} instead of the integral
expressions~\Ref{10j} and~\Ref{10angles}. The result is an expression
for the correlations $W$ with no integrals, but a large number of sums.
Until recently, this spin summation method has been intractable due to
the large number of operations needed to evaluate the sums. However, the
techniques developed in~\cite{Khavkine} have made
it feasible and even advantageous to the Monte Carlo method in a certain range
of $j_0$. The improvement comes from taking advantage of the
factorization of the boundary state~\Ref{psiFact} and of hidden
structures in the sum to significantly reduce the number of operations
needed to evaluate it. A more detailed outline of this efficient spin
summation method is given in the appendix.

Both the Monte Carlo and the spin summation methods are approximate, the
former requiring more samples for better convergence and the latter
requiring larger truncations of an otherwise infinite sum for the same
effect (essentially, restricting the boundary spins to lie within a
finite interval about $j_0$). The difference is that, for
$j_0$ for which the spin summation method is feasible, its convergence
is much sharper than that of the Monte Carlo method. In that range, spin summation gives
at least 5 decimal places of accuracy compared to 2 or fewer
for a Monte Carlo run of similar length.
Fortunately, the feasibility
of the spin summation method extends over a large enough range of $j_0$
and is of sufficient precision to allow us to probe the next-to-leading
order behaviour of the correlations $W$ as a function of $j_0$.

\subsection{Numerical results}

We study the area correlator \Ref{W1} as a function of the squared distance $j_0$ between the two areas, with
attention to how it is affected by changes in the boundary state, i.e.\ by changes to the parameters $\alpha$ and $k$.
First of all, notice that thanks to the symmetries of the background, there are only three independent
quantities in \Ref{W1}, which correspond to $a$ and $b$ being (i) the same triangle, (ii) two adjacent triangles,
(iii) two opposite triangles.
Here we report the data for case (iii). 
The plots for the other components are qualitatively the same.

\begin{figure}[bt]
\includegraphics[width=7.5cm]{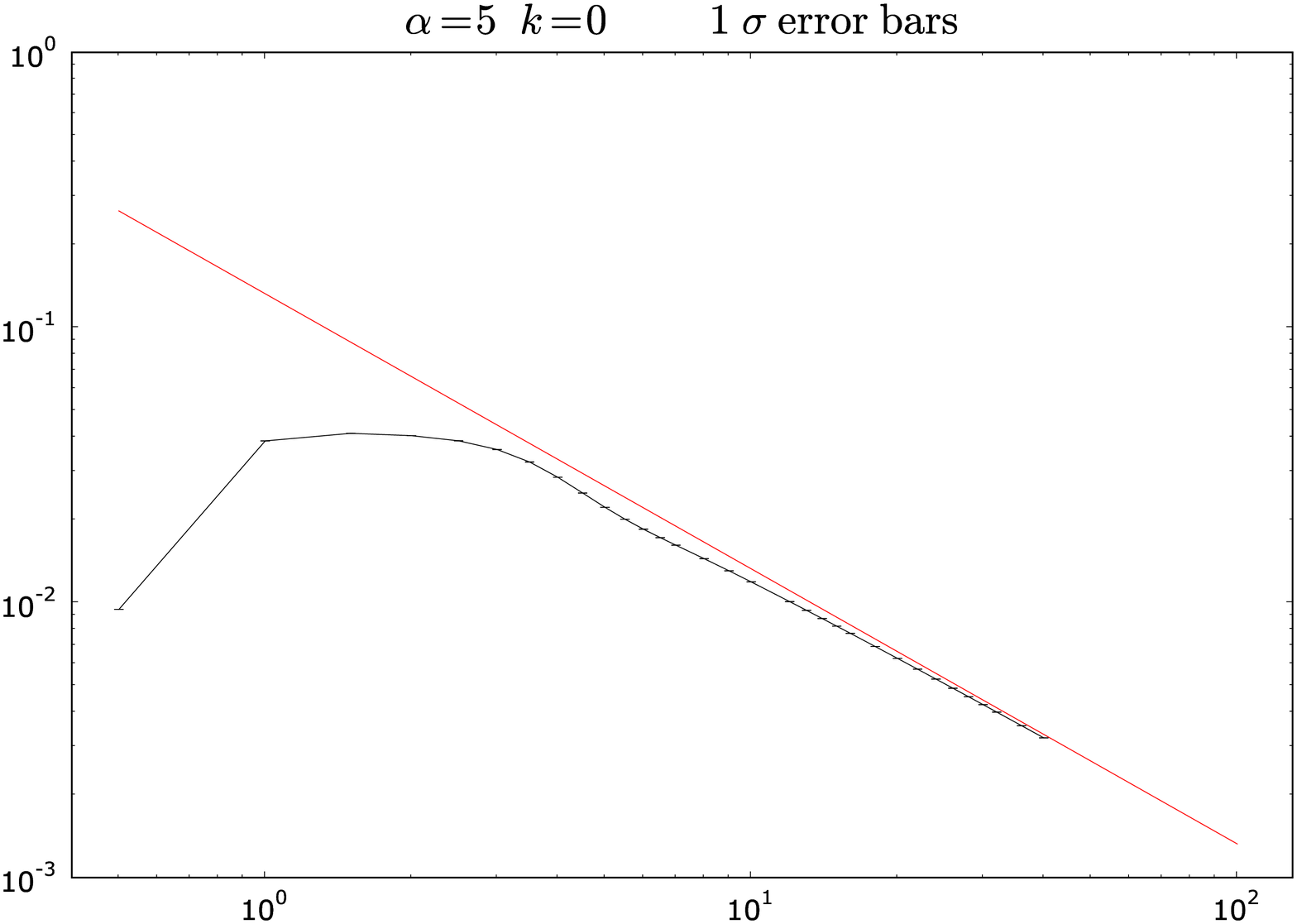}
\includegraphics[width=7.5cm]{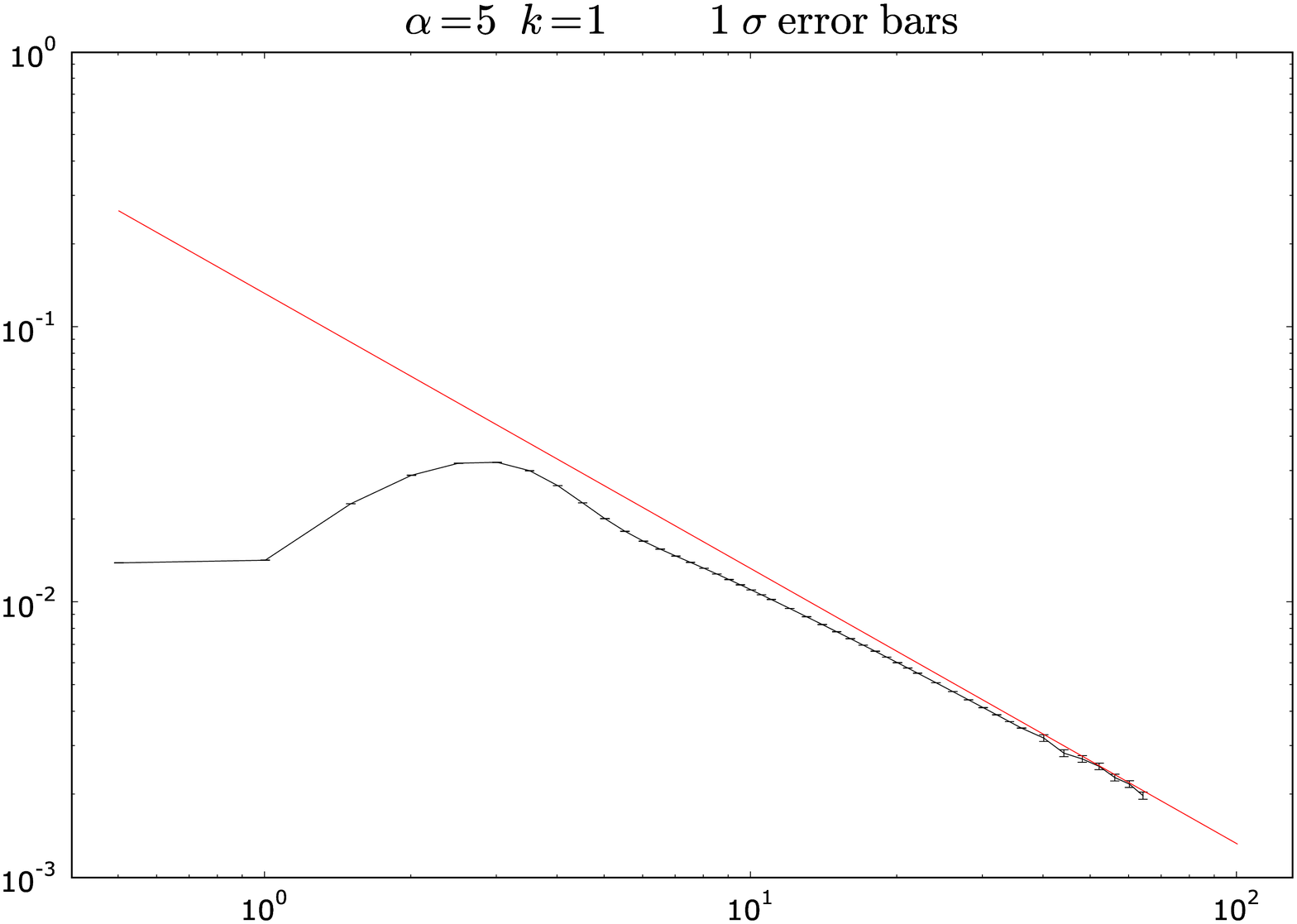}
\includegraphics[width=7.5cm]{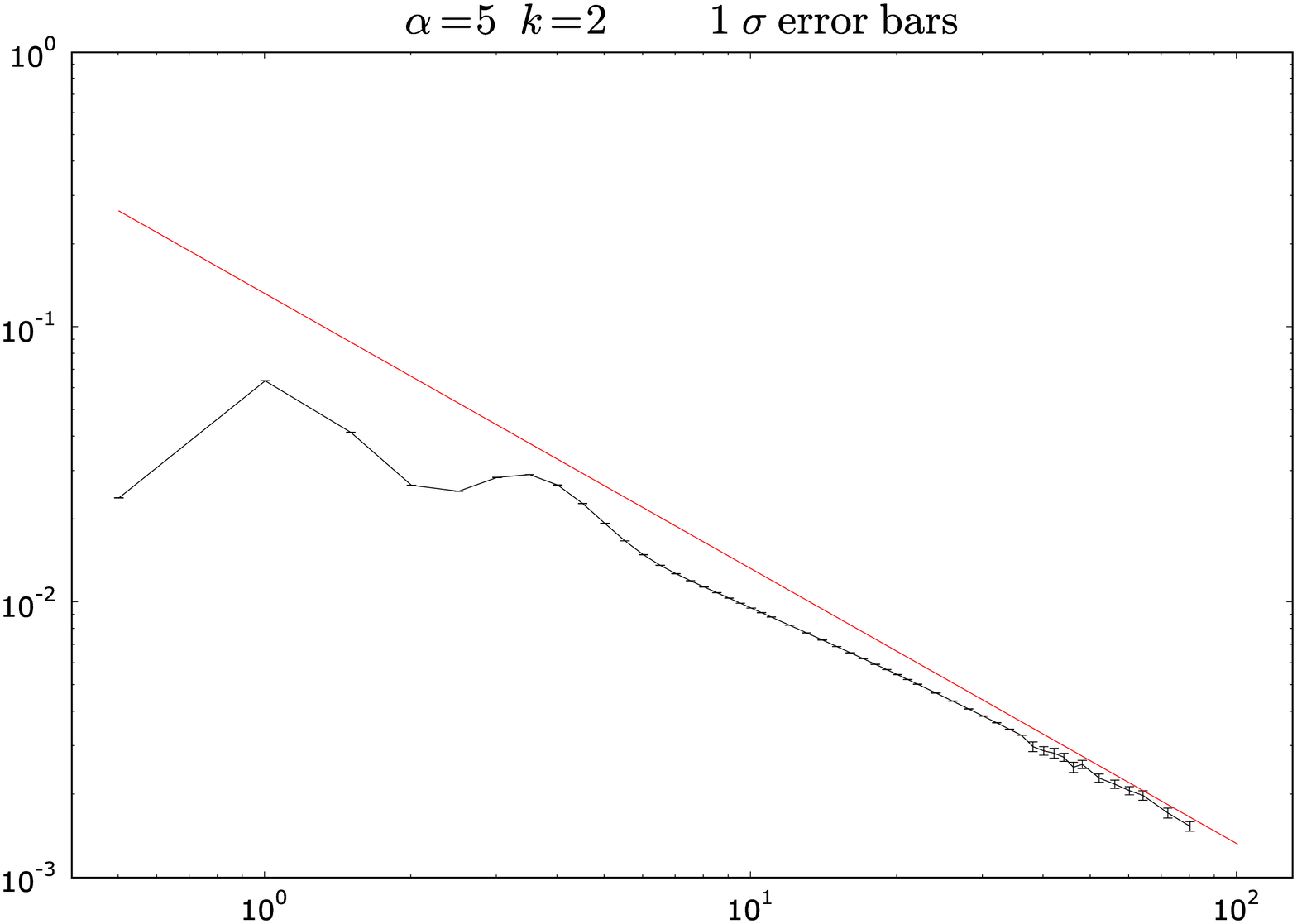}
\includegraphics[width=7.5cm]{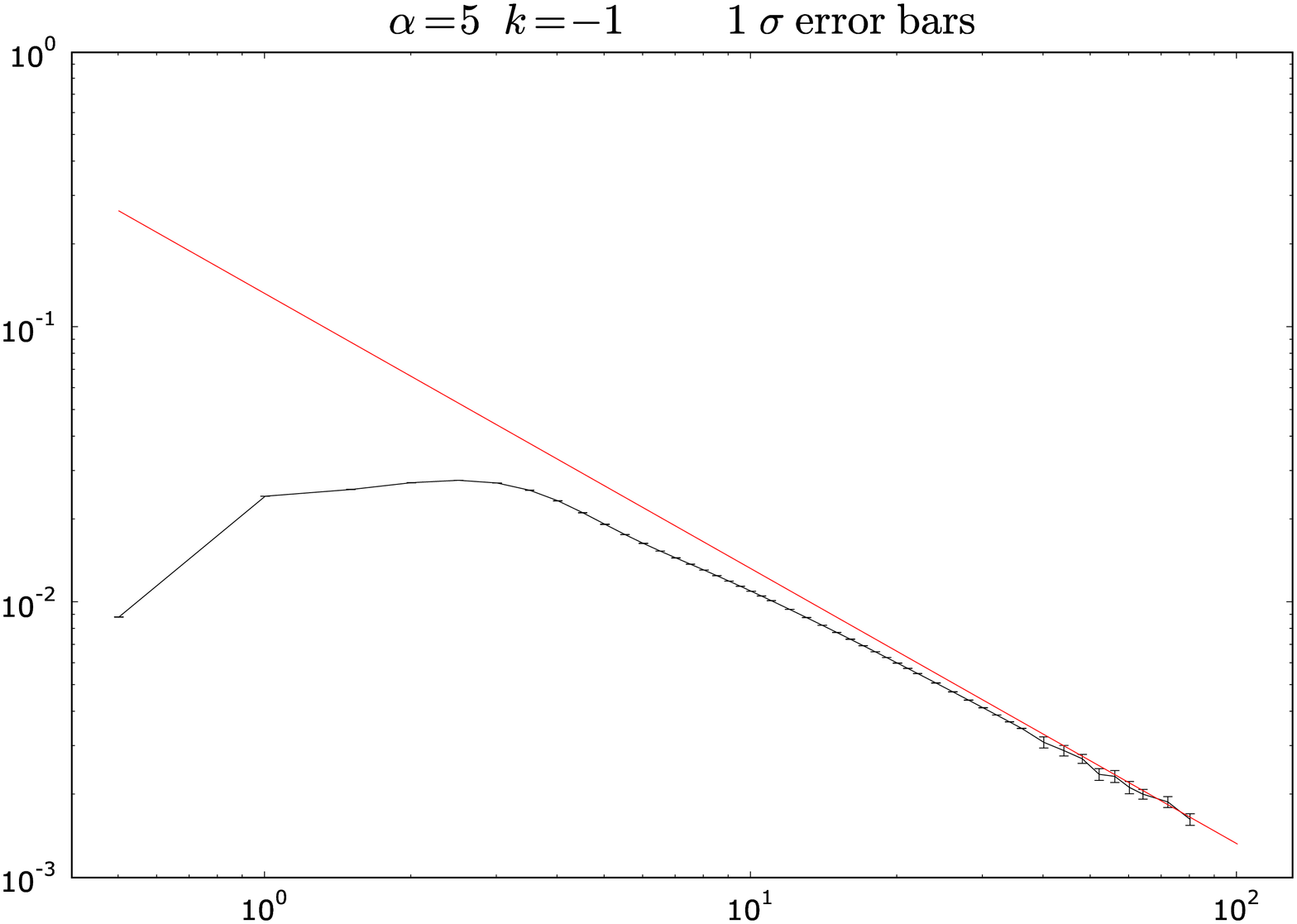}
\caption{\small{Log-log plots of a component of \Ref{W1} (opposite triangles) for $\alpha=5$ and $k=0,1,2,-1$ as a function of $j_{0}$. The interpolated dots are the numerical evaluation, with error bars in some cases too small to see, and the lines the analytic calculations of the leading order.}}\label{FigExact}
\end{figure}

\begin{figure}[bht]
\includegraphics[width=7.8cm]{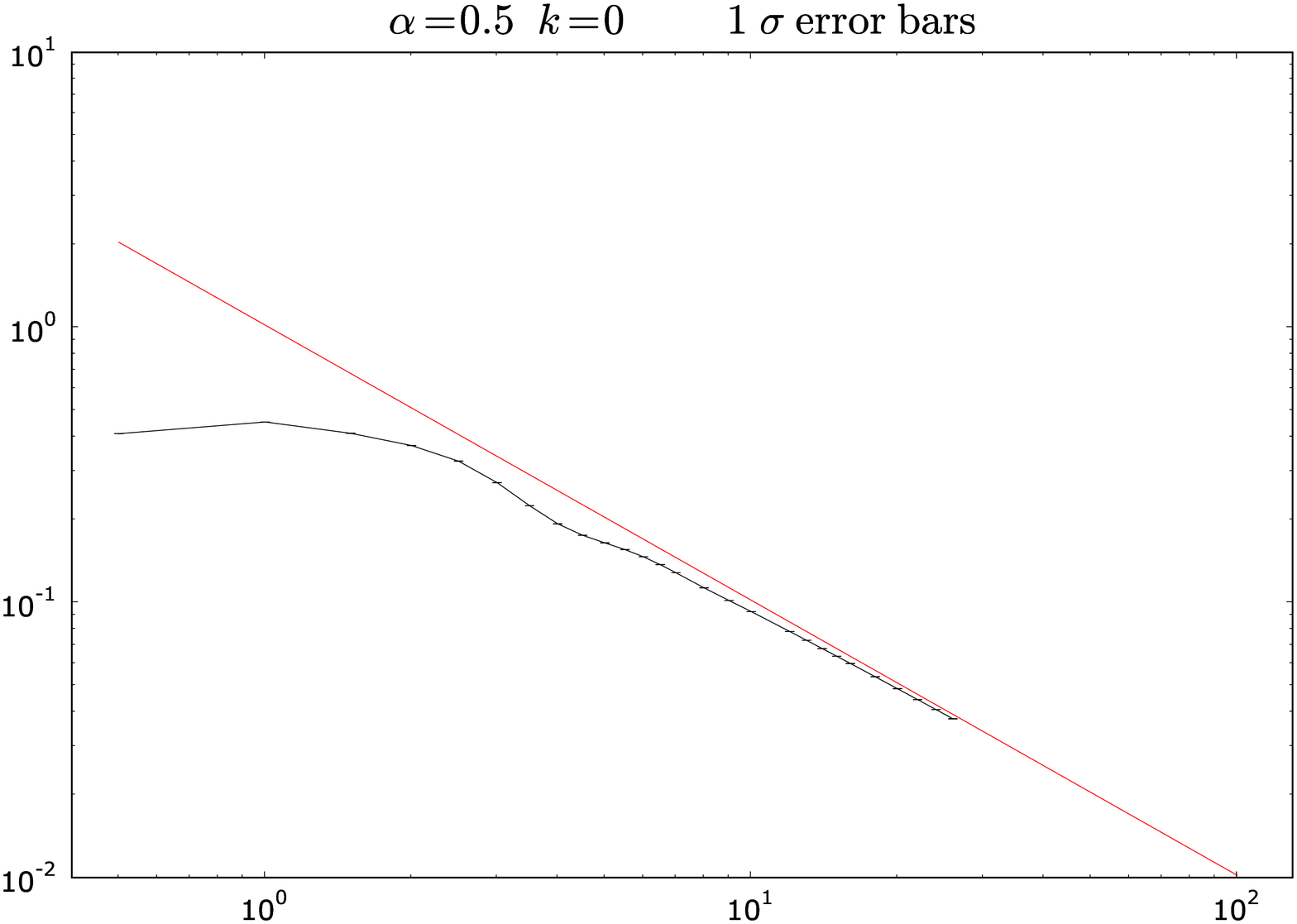}
\includegraphics[width=7.8cm]{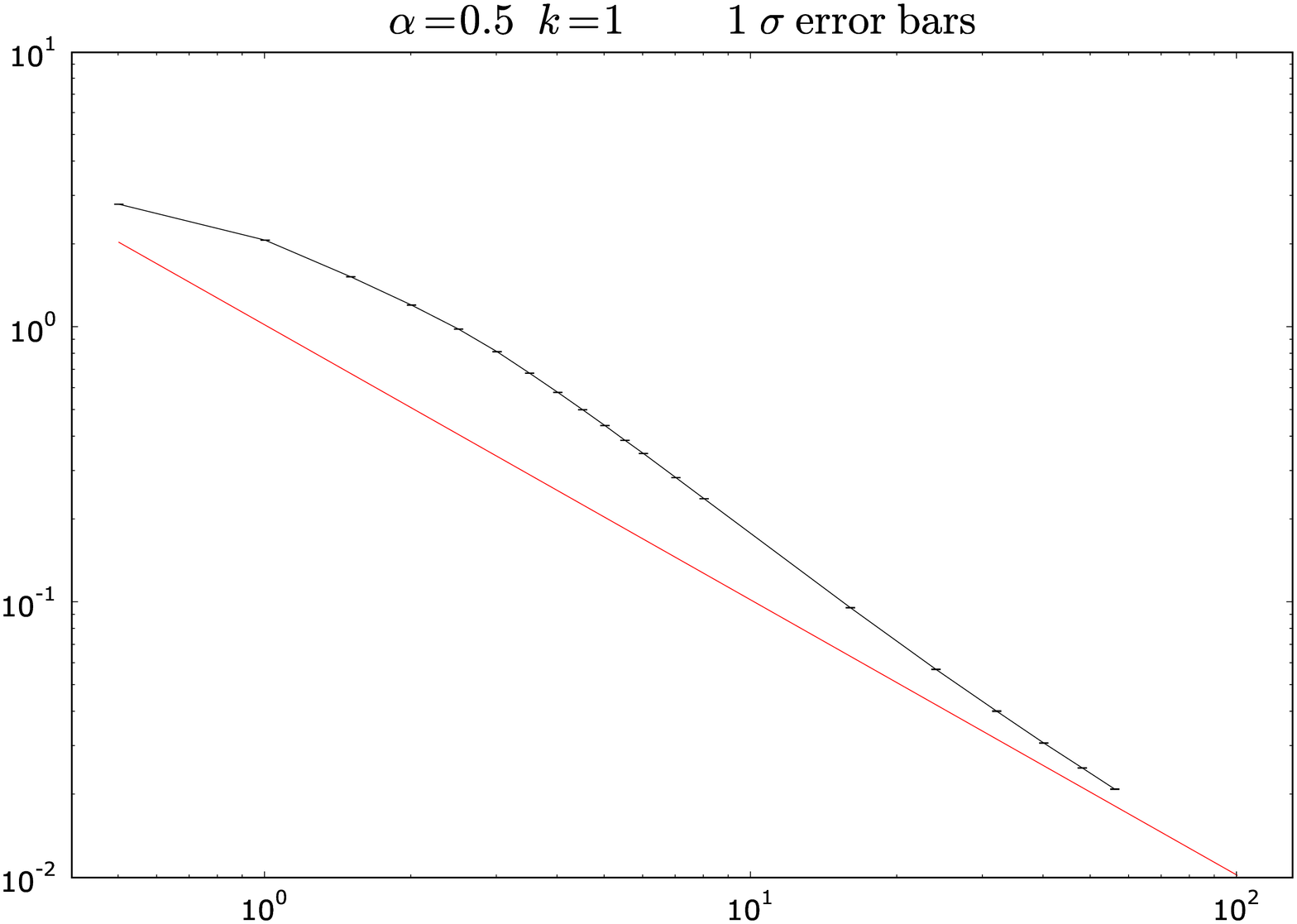}
\includegraphics[width=7.8cm]{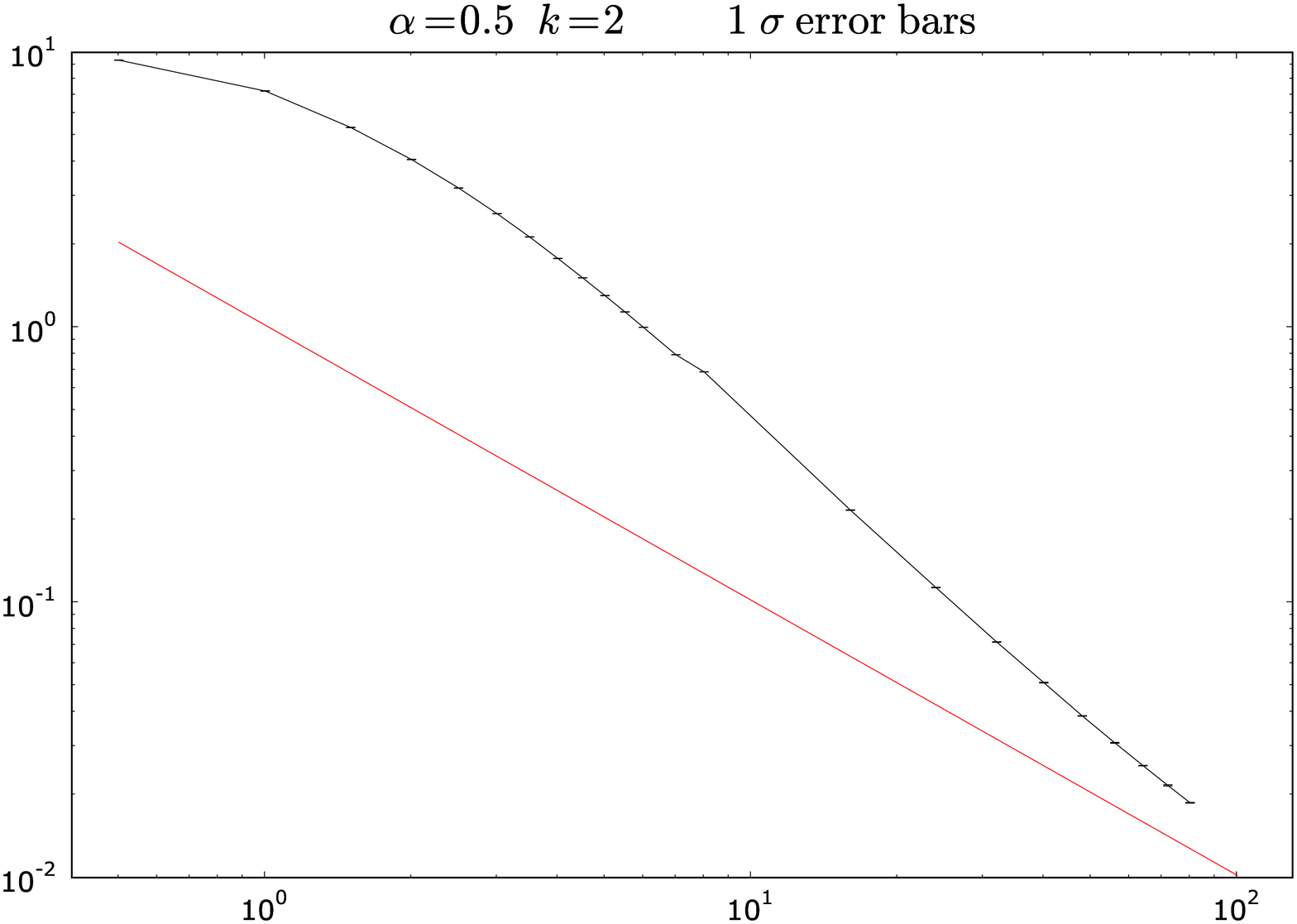}
\includegraphics[width=7.8cm]{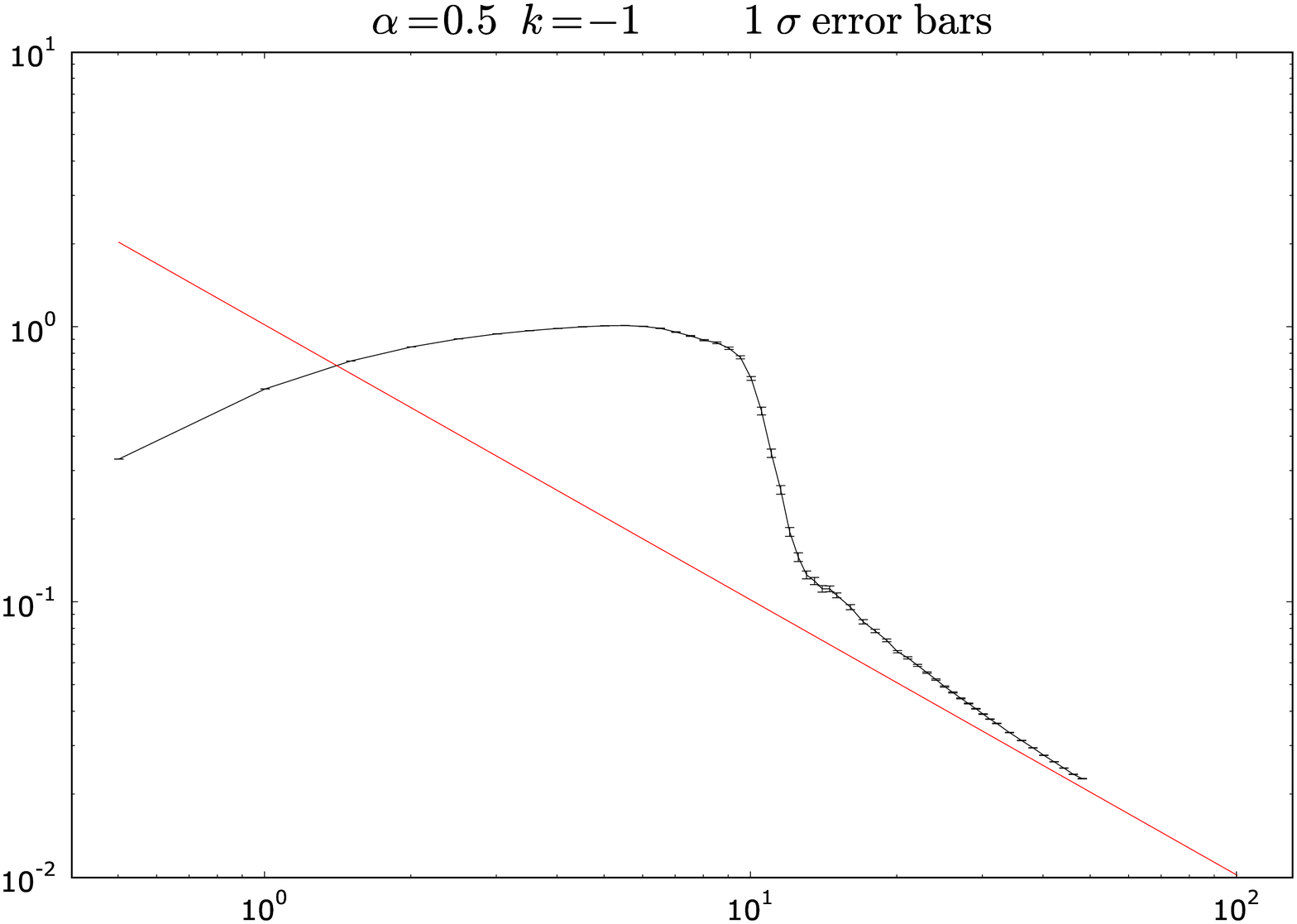}
\caption{\small{Log-log plots of the same component for $\alpha=0.5$ and $k=0,1,2,-1$ as a function of $j_{0}$.}}\label{FigExact1}
\end{figure}

A brief overview of the numerical studies is presented in Figures \ref{FigExact} and \ref{FigExact1}, for the cases $\alpha=5$ and $\alpha=0.5$, and different choices of $k$.
%
The plots show that the leading order behaviour is proportional to
$1/j_0$, with coefficient depending on $\alpha$. This is compared with the analytic calculations, performed following the technique of \cite{RovelliProp}, finding accurate agreement as already reported in \cite{letter}. We refer to the latter for details on the analytic calculations.
We stress the fact that measure terms like the $k$-dependent triangle amplitudes do not affect the asymptotic leading order, in agreement with \cite{grav3,3d} and intuitively to be expected, since ambiguities in the path integral measure should not affect the leading semi-classical behaviour, but only the quantum corrections. Indeed, the triangle amplitudes do change the corrections and affect the way the asymptotic regime is reached. For instance, it is reached faster for $k=0$,
hinting that this choice minimizes the quantum corrections. We will look at this in more details in the next section.

The plots also show the short scale picture where the non-perturbative regime is probed. As presented in \cite{letter}, the surprising feature is the existence of a maximum in the correlations, followed by an inversion of the scaling behaviour:
as we go to smaller $j_0$, we reach a peak and then the correlations do not diverge but instead are completely regular for small values of $j_0$. The precise position of the peak depends on the value of the (Gaussian) width $\alpha$ and on $k$.
We will come back to this point in more detail below in Section \ref{sectionpeak}.
Finally, a closer look at the plots shows also some residual oscillations. These are most likely due to the higher order terms in the expansion of the Regge action around the background, which contribute an oscillatory behaviour as they appear in complex exponentials (see e.g.\ \cite{RovelliProp} for details).

\section{Perturbative expansion}
\label{nlosection}

To investigate more quantitatively how the measure term affects the way the asymptotic regime is reached,
we fit the leading and next-to-leading orders of the numerical evaluations. This is a place where the techniques from~\cite{Khavkine}, adapted to the present situation, have been crucial. Computing the graviton propagator using summations over spins gives sufficient precision to examine next-to-leading order asymptotic behaviour, unlike the purely numerical (Monte-Carlo) integration over group variables.

After fitting power laws and numerical coefficients, we find that the correlations have the following structure:
\equ\label{W2}
W_{ab}(k, \alpha, j_0) = \f{f^{(1)}_{ab}(k,\alpha)}{j_0}+  \f{f^{(2)}_{ab}(k,\alpha)}{j_0^2} + o(j_0^{-2}).
\nequ
Some explicit fits are reported in Table 1.

\begin{table}[htb]
	\centering
	\begin{tabular}{|c||c|c|c|c|c|}
	\hline
			$(\alpha, k)$ & (5, -1) & (5, 0) & (5, 1) & (5, 2) & (0.5, 0) \\  \hline
			$|f^{(1)}_{\rm opp}|$ & 0.127 & 0.130 & 0.126 & 0.120 & 0.982 \\ \hline
			$|f^{(2)}_{\rm opp}|$ & 0.249 & 0.149 & 0.241 & 0.499 & 0.960 \\
        \hline
		\end{tabular}\caption{Numerical fits of the LO and NLO for some values of $\alpha$ and $k$.	 Only the cases when $a$ and $b$ are opposite triangles are reported. 
}
\end{table}
After the leading order (LO) in $1/j_0$, the next-to-leading order (NLO)
is in $1/j_0^{2}$. This linear power increase is natural given the asymptotic nature of the
perturbative expansion. The NLO in large $j_0$ can in fact also be computed analytically,
using for instance a saddle point approximation as in \cite{grav3}. However, calculating higher order corrections becomes quickly very lengthy and cumbersome, due to the high number of different contributions, already for a single 4-simplex. To get an idea, compare the complexity already reached by analogous three-dimensional calculations on a single tetrahedron
 \cite{3d2}, and even just for the underlying $6j$-symbol \cite{maite1}. Numerical methods allow us to focus directly on the global structure of the perturbative expansion.

The fitted values for $f^{(1)}$ reported in the Table approximate well the analytically computed values \cite{letter},
\equ
f^{(1)}(\alpha=5) = 0.133, \qquad f^{(1)}(\alpha=0.5) = 1.02.
\nequ
From the Table we can observe a slight mismatch on the second digit of the $f^{(1)}$ for different $k$. This can be reduced by pushing the simulations to higher $j_0$, however this would be computationally very expensive, and it is already sufficiently reassuring to see the one digit match, together with the plots in Figure \ref{FigExact} showing clearly that the predicted line is being approached. The slight mismatch between the  coefficients in the table also reflects the fact that the speed with which we reach the asymptotic regime depends on the value of $k$.

\begin{figure}[ht]
\includegraphics[width=6cm]{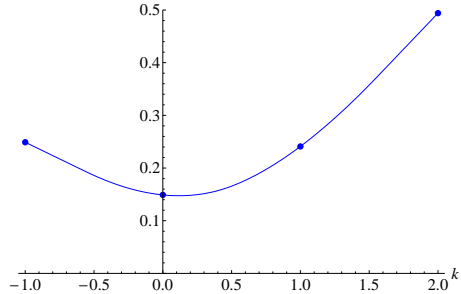}
\caption{\small{The blue dots are the values of $f^{(2)}(\alpha=5)$ at different $k$.
The interpolating line is not a fit, it is a simple line linking the plotted values.}}\label{FigNLO}
\end{figure}

Let us now focus attention on the next-to-leading order. The table shows that the NLO depends upon $\alpha$, just like the LO, but more interestingly it shows a significant dependence on $k$. To highlight this dependence, we plot in Figure \ref{FigNLO} the values of $f^{(2)}(\alpha=5)$ as a function of $k$. The plot shows a roughly quadratic dependence with a minimum near the value $k=0$.
This result can be compared with the three-dimensional case studied in \cite{3d}, where a similar quadratic dependence of the NLO coefficient was found and shown analytically.\footnote{For the reader familiar with the perturbative expansion of the three-dimensional geometrical correlations introduced in \cite{Io} and further studied in \cite{3d,3d2}, we would like to point out an interesting difference. Recall that both in three and four dimensions, the origin of the good behaviour of the leading order comes from the emergence of a Regge-like action from the spin foam kernel, the Wigner $6j$-symbol in 3d and the $10j$-symbol in 4d. In \cite{3d}, we distinguished two sources of quantum corrections to the leading order: the ones coming from higher
orders of the Regge path integral, called (i) in the reference, and the ones coming
from higher orders of the expansion of the $6j$-symbol, called (ii). In particular, in \cite{3d} it was found
that only (i) contribute to the next-to-leading order on a single tetrahedron, whereas (ii) entered at
next-to-next-to-leading order. A similar situation might arise in the current case, although we are not in a position to make a precise statement
since there has not been any computation of the higher orders of the expansion of the $10j$-symbol. A difference that can be
pointed out is that the situation could be even richer, due to the degenerate non-Regge-like terms discussed previously.}
It would be interesting (and indeed possible, with enough courage) to compute analytically the NLO in order to see the exact dependence.

What can we read from the form \Ref{W2} of the perturbative expansion?
Recall the standard perturbative calculation of the Newton force \cite{Bjerrum},
\equ\label{VN}
F(r)= G \f{m_1 m_2}{r^2} \left( 1 + a_1 G \f{m_1+m_2}{r} + a_2 \f{\hbar G}{r^2} \right).
\nequ
The precise value of the numerical coefficients $a_i$ does not matter for our considerations. The term proportional to $a_1$
is a classical correction from general relativity, whereas $a_2$ is the first quantum correction.
The universality (i.e.\ mass-independence) of the quantum correction allows one to define a running
Newton's constant
\equ
G(r) = G \left(1 + a_2 \f{\hbar G}{r^2} \right).
\nequ
The result \Ref{VN} is computed
coupling the graviton propagator to static external matter sources. At the state of the art of the formalism, we are far from being able to do this: we need to add matter, deal with more than a
single 4-simplex, plus have a fully dynamical boundary state. Nevertheless, it
is interesting to look at the different powers of $r$ and attempt a match with \Ref{W2}.
The leading order behaviour in $1/j_0$ is consistent with the
$1/r^2$ scaling of the force in \Ref{VN}. 
Next, the NLO in $1/j_0^2$ points directly at the first quantum correction in \Ref{VN}, so again we might say that we obtain a consistent scaling of the NLO.
Of course, the reader can immediately notice the absence of a term in $1/j_0^{3/2}$ in \Ref{W2},
which means the absence of the classical correction, i.e.\ the term in $a_1$ of \Ref{VN}. The simplest explanation is that one needs to couple matter to see this term explicitly, so its absence should not be taken as further evidence of the inappropriateness of the low energy dynamics encoded in the Barrett-Crane model; indeed, the same absence could occur when considering the new EPR-FK-LS spin foam models if matter is not coupled.

Since the corrections to the Newton potential are in principle observable, the discussion above shows that fixing uniquely the face amplitudes (in the path integral measure) is one of the necessary steps to make a spin foam model predictive.

\section{Short scale behaviour and metric renormalization}
\label{sectionpeak}

\begin{figure}[htb]
\includegraphics[width=7.9cm]{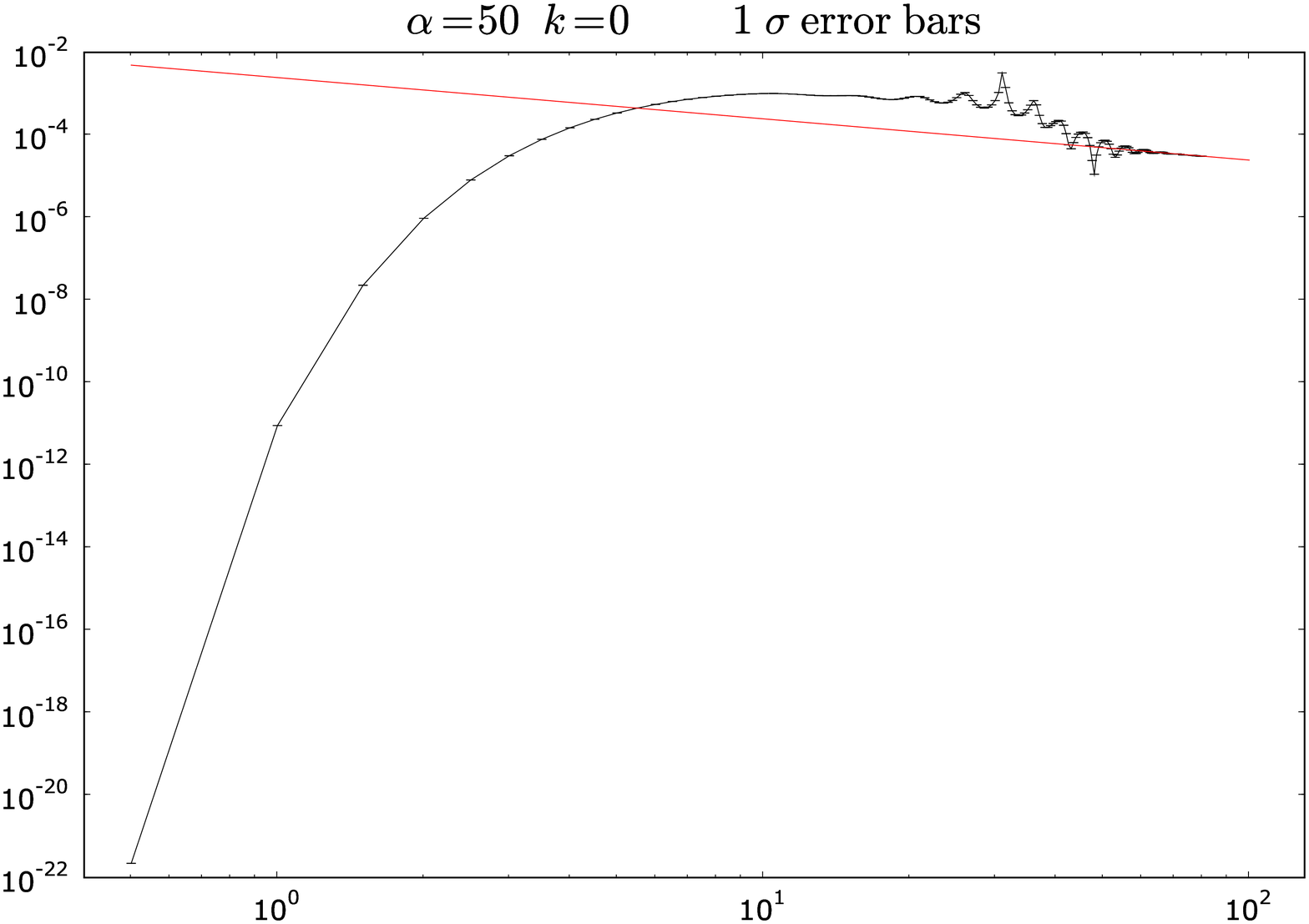}
\includegraphics[width=7.9cm]{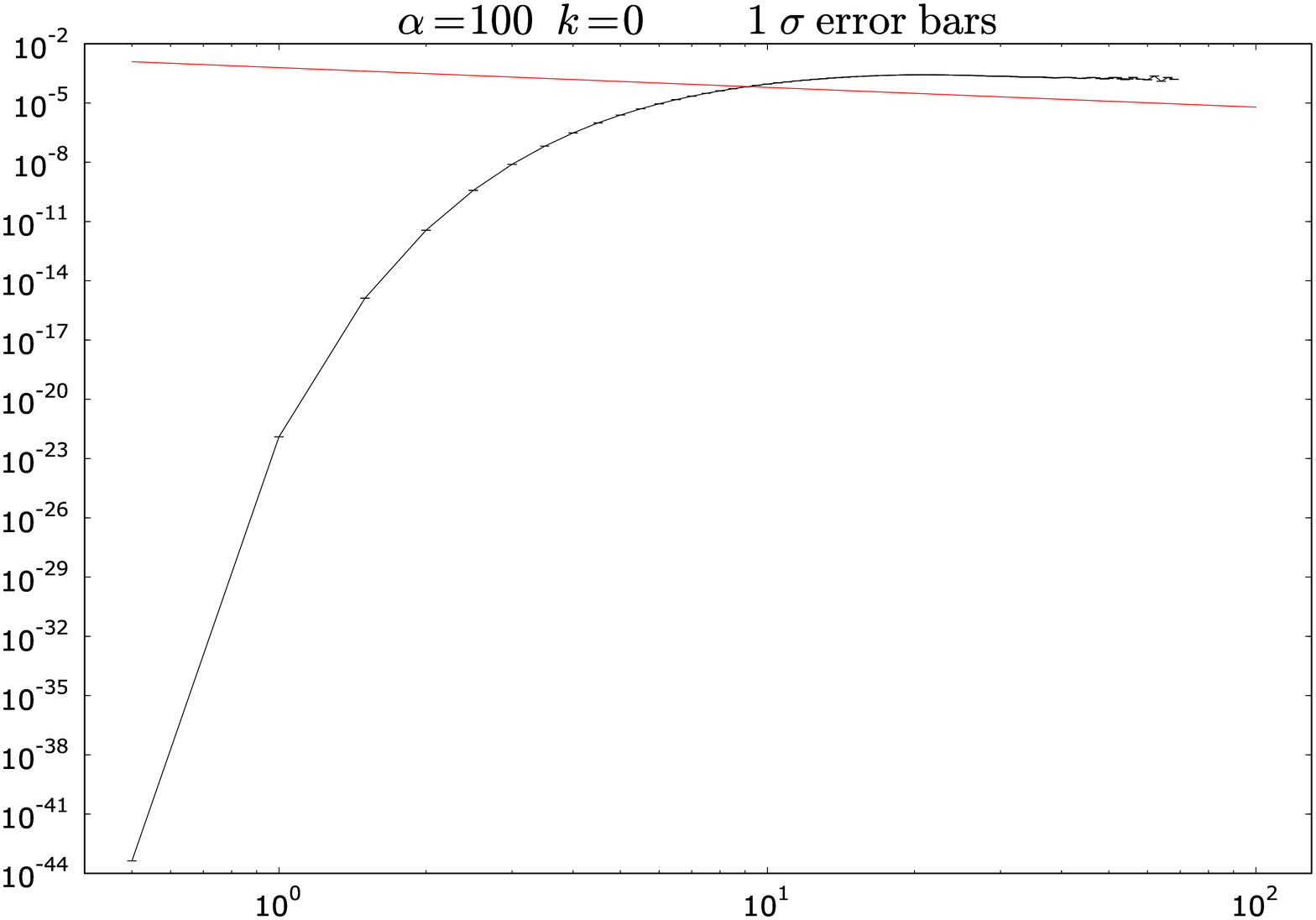}
\caption{\small{Log-log plots for $\alpha=50$ and $\alpha=100$, $k=0$, as a function of $j_{0}$. The lines are the analytic calculations of the leading order. The (first) peak is at $j_0\sim o(10)$. Notice the presence of oscillations at intermediate scales.}}\label{FigExact2}
\end{figure}

The numerical simulations not only allow us to confirm the leading order and study the $k$-dependence of the NLO,
but also offer us a look at the possible non-perturbative effects of spin foam quantum gravity at short scales,
albeit in the simplified model here considered. As already presented in the short paper \cite{letter}, the key point is that, instead of keeping the usual inverse power behaviour, the correlations at short scale are regular and do not diverge; they reach a peak at some multiple of the Planck scale, after which the correlation recovers its standard behaviour in the inverse squared distance. The exact position and shape of the peak depend on both $\alpha$ and $k$, see Figs.\  \ref{FigExact} and \ref{FigExact1}.
The peak moves (very slowly) to larger spins as we increase the value of $\alpha$, cf.\ Fig.\ \ref{FigExact2} for the cases $\alpha=50$ and $\alpha=100$: the location of the peak is now at values of $j_0$ of order 10.

These plots also show some unexpected oscillations at an intermediate scale. The frequency of these oscillations suggests that they are related to higher order terms in the expansion around the background of the Regge action in \Ref{10jasymp}, however we are not able at this stage to give a more precise explanation. Let us nevertheless point out that such oscillations already appeared in the three-dimensional case  for intermediate spins (when looking at the NLO) \cite{3d}. In that case, it concerned residual oscillations which come from the phase of both the spin foam amplitude and the boundary state, and they could be tamed by changing the width of the Gaussian.
We do not discuss these oscillations further and instead focus on the short scale picture.

How can we interpret the presence of a peak, and the very unusual behaviour of the correlations at scale shorter than the peak?
In a truly background independent formalism, the semiclassical geometry $q$ is a low-energy property of a quantum state $\ket{q}$, which defines the boundary state $\Psi_q[g_{ab}]=\langle{g_{ab}}\ket{q}$, where $a$ and $b$ are three-dimensional indices.
It is natural to expect that the expectation value of the bulk geometry $g_{\mu\nu}$ depends upon the
energy scale set by the boundary, i.e.\ $j_0$.
The correlations then allow us to reconstruct the effective metric.
It is indeed natural in a background independent context to reconstruct the space-time geometry and
metric from the only available information, namely the correlations (see e.g.\ \cite{danny}). Thus
we distinguish the flat metric $\eta$ and the actual (effective)  metric $g$. Assuming that the
correlations $W_{ab}$ scale as $1/d_g(x,y)^2$ in terms of the effective metric, the correlation
graph gives the evolution of $g$ in terms of the scale $d_\eta(x,y)\sim \sqrt{j_0}$ (remember that $j_0$ is the area scale, thus the square of the length scale). In the asymptotic
regime for large $j_0$, the metric $g$ converges towards the flat metric $\eta$ and we have the right semi-classical limit. For small distances $d_\eta(x,y)$, we have strong quantum gravity effects and the effective metric $g$ deviates from $\eta$.

From this point of view, the correlation peak actually provides a dynamical
definition of a minimal distance: the physical distance $d_g(x,y)$, in term of the effective
metric, is never smaller than $j_0^{min}$.
This minimal scale result can be compared to similar singularity avoidance results obtained in the framework of Loop Quantum Cosmology (see e.g.\ \cite{lqc}).

This dynamical minimal scale $j_0^{min}$ depends on the choice of
boundary state. It indeed depends on the precise value of $\alpha$. Changing the boundary state
would modify the bulk geometry and the value of this transition scale between semi-classical
behaviour and quantum gravity regime. However, we hope that the boundary state and the value of the
Gaussian width $\alpha$ will be fixed by the requirement of working with a physical state \cite{Bianca,PhysBS}. This would lead to a precise prediction for the minimal
scale\footnotemark~$j_0^{min}$.

Actually, at scales shorter than the maximal correlation peak at $j_0^{min}$, the physical distance
$d_g(x,y)$ increases as $d_\eta(x,y)$ gets smaller. This is a sign that the metric becomes more
and more curved as we get to shorter scales. This is consistent with the expected picture that the
quantum fluctuations of the geometry create high curvature (and possibly virtual black holes) at
the near-Planck scale. We recall that a correlation growing as the one we see is a sign of a
potential which grows with the distance and thus of confining effects.

\footnotetext{
Here, we have been working with the Riemannian theory and a compact gauge group $\Spin(4)$. The
representation labels $j$ are then discrete. Therefore, we already have a notion of minimal length scale
given by the smallest non-trivial representation $j=1/2$. Notice that our notion of dynamical minimal length is different from this. 
In the Lorentzian theory \cite{BC2}, the representation labels become continuous. Assuming that the correlations behave similarly in that case, our definition would lead to a non-trivial notion of
dynamical minimal length scale. Thus, it would very interesting to numerically check the Lorentzian model to see if it has the same short scale behaviour for the correlations.}

In this respect, let us recall that different approaches to quantum gravity have recently seen evidence that the structure of space-time becomes effectively two-dimensional at short scales.\footnote{This has been seen in Causal Dynamical Triangulations \cite{CDT} and in the Exact Renormalization Group approach \cite{erge} based on the idea of asymptotic safety \cite{Weinberg}. See also \cite{Benedetti} for possible similar results in the context of quantum group models of quantum gravity, and \cite{Modesto} for a tentative extension of these ideas to loop quantum gravity.}
In terms of the area correlations considered here, such a result would show up if the
$1/j_0$ large scale behaviour of the correlations is replaced at short scales by $\log j_0$.
This is not precisely what we see, but rather a roughly linear growth. This bears some qualitative similarities with possible confinement properties of the theory, but of course any such speculation should be postponed to more realistic models. In particular, to obtain the exact effective metric, we need the full tensorial structure of the
propagator, which can only be obtained using the more complete new models.

Nevertheless this is a crucial line of investigations, and our results show that numerically studying the geometric correlations is a fundamental tool for assessing the strength of the spin foam formalism to address the long-standing questions of quantum gravity.

\section{The Bessel-based boundary state}
\label{besselsection}

\begin{figure}[ht]
\includegraphics[width=7.5cm]{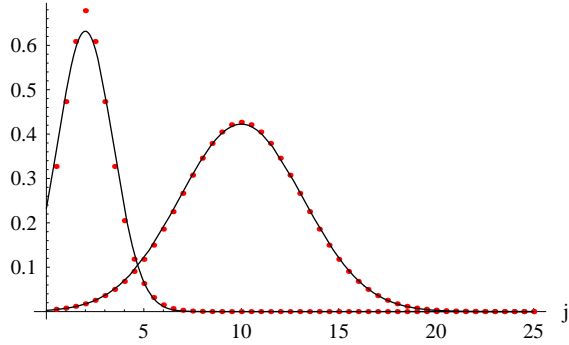}
\caption{\small{Comparison of the Bessel state \Ref{psiB} (dots) with the Gaussian state \Ref{psiG} (solid lines),
for $j_0=2$ and $j_0=10$.}}\label{BvsG}
\end{figure}

To complete the overview of our numerical studies, we report in this section the results obtained
using the Bessel-based boundary state \Ref{psiB}.
Although very reasonable in the large spin limit, the Gaussian ansatz can not be
considered an exact state in the full theory: the conjugation between the dihedral angle and the spin is only
achieved taking the latter as real variables, an assignment that can be justified only for large spins.
On the other hand, in the full theory the $j_l$ are half-integers and are conjugated to the $\SU(2)$ class angles
(see for instance \Ref{10j}) in the sense of harmonic analysis.
It thus appears natural to look for a boundary state where the conjugation between the spins and the dihedral angles is realized through harmonic analysis over $\SU(2)$. This is achieved using
\Ref{psiB}. This state was introduced in \cite{3d} for the three-dimensional theory and extended
to the four-dimensional theory in \cite{grav3}.
The large spin approximation of the Bessel functions in \Ref{psiB} gives precisely the Gaussian appearing
in \Ref{psiG} \cite{3d},
\equ
\f{ I_{|j-j_0|}(\f{j_0}{\alpha})-I_{j+j_0+1}(\f{j_0}{\alpha})}
{\sqrt{ I_{0}(\f{2j_0}{\alpha}) - I_{2j_0+1}(\f{2j_0}{\alpha})}}
\simeq \sqrt[4]{\f\alpha{{j_0}\pi}}\exp\{-\f\alpha{2{j_0}}(j-j_0)^2\}.
\nequ
This asymptotics is reached very early and with great accuracy, as can be seen
in Figure \ref{BvsG}.
Therefore this choice of boundary state gives the same leading order of the area correlations, and also
similar corrections and short scale picture, provided we take the same phases.
On the other hand, there is an advantage if we change the phase to the cosine given in \Ref{psiB}, as in this case we can perform the sums \Ref{w} \emph{analytically} \cite{grav3}: restricting $|j-j_0|\in {\mathbbm N}$, we have \cite{3d}
\equ\label{FT}
\widetilde\psi_{\scr B}(g)=\sum_j \psi_{\scr B}(j)\, \chi_j(g)= \sum_{\eta=\pm}
\f{e^{-\f{2j_0}{\alpha}\sin^2(\phi-\eta\theta)}}{2e^{-\f{j_0}{\alpha}}N\sin\phi}\sin \Big(d_{j_0}(\phi-\eta\theta)\Big).
\nequ
Thanks to this expression, one can now evaluate \Ref{w} analytically for $k=0$.\footnote{One can also
treat in this way all $k$ even cases, exploiting the property
$(2j+1)^{2n} \chi^j(\phi)=\n_\phi^2  \chi^j(\phi)$. Generalizations to $k$ odd are on the other hand less obvious.
}
The resulting expression for the correlator \Ref{W1} takes the form
\equ\label{WB}
W_{ab}(k,\alpha,j_0)\,=\,  \f{1}{{\cal N}} \sum_{\{\eta_l=\pm\}} \int \prod_l d\phi_l\, C[\phi_l]\,
\wtl{w}_{\eta_a}(\phi_a)\wtl{w}_{\eta_b}(\phi_b)\prod_{l\ne a,b} w_{\eta_l}(\phi_l).
\nequ
This was introduced and studied in \cite{grav3}, to which we refer for details.
Obtaining an exact integral representation for \Ref{W} was the initial rationale to introduce the new boundary state
in \cite{grav3}, precisely with the hope of improving analytic and numerical control of \Ref{W}. From the analytic point of view, the integral representation was instrumental in proving the conjecture of \cite{RovelliProp} that the degenerate terms of the Barrett-Crane asymptotics do not contribute to the leading order. The use of this state was also important from the numerical point of view, as numerics for \Ref{psiB} where the first ones to be obtained, confirming the analytic calculations of \cite{grav3}, and paving the way to the numerical procedure later used for the Gaussian state, and presented in \cite{letter} and here.

\begin{figure}[ht]
\includegraphics[width=7.8cm]{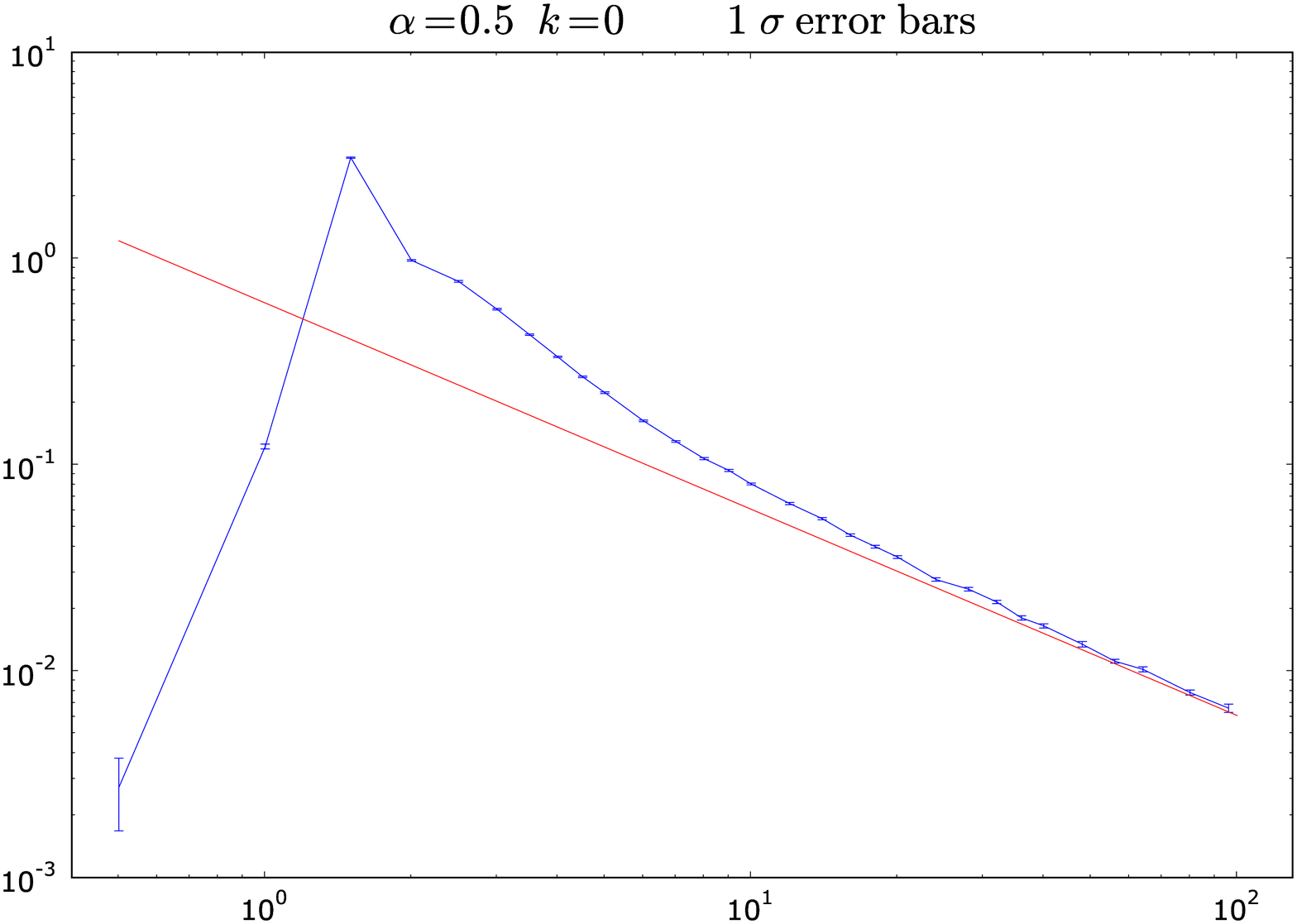}
\includegraphics[width=7.8cm]{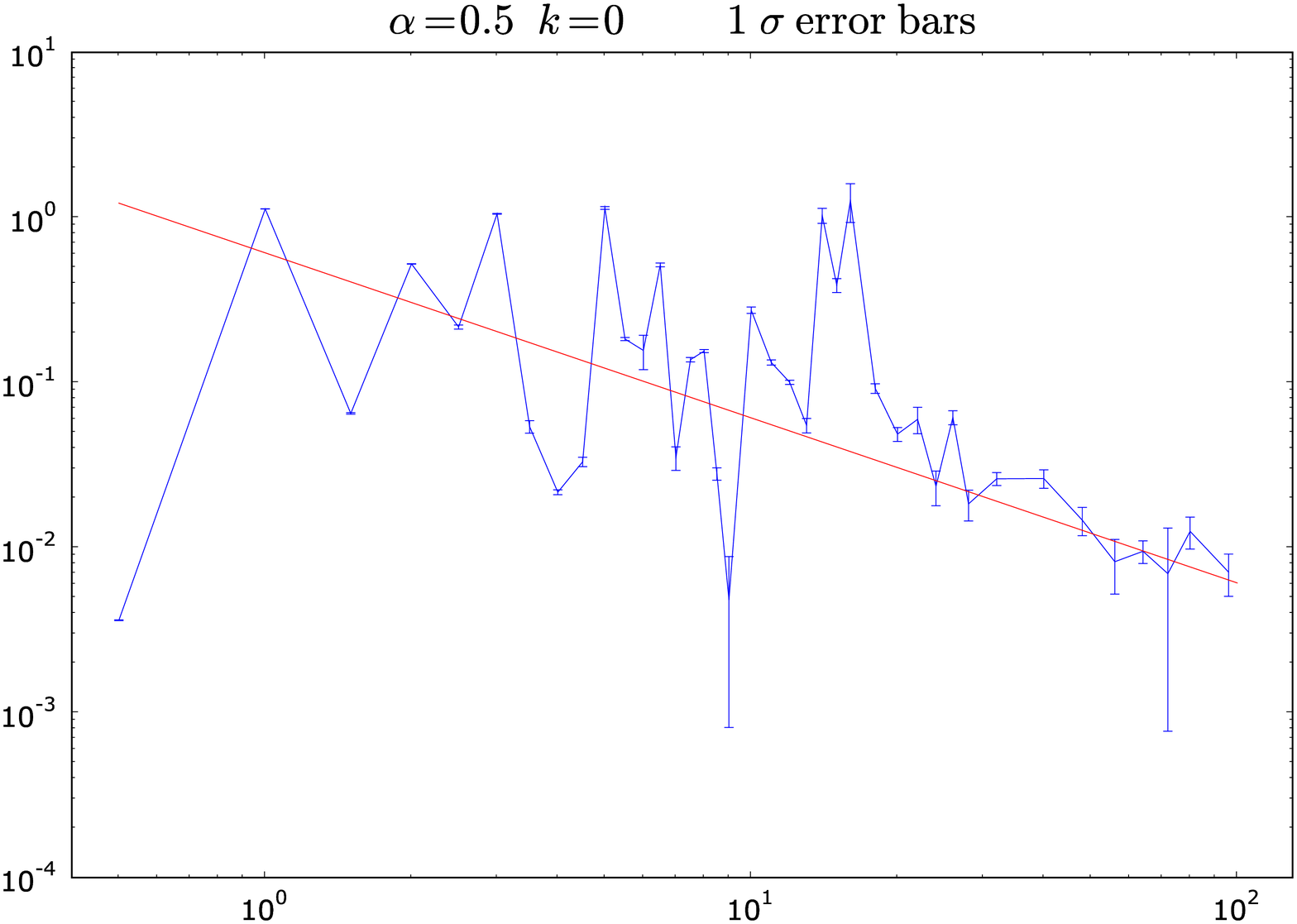}
\caption{\small{The Bessel plots. Left panel: the coherent signs configuration, namely the single $\eta_l=1,\, \forall l$. Right panel: the result after summing over the etas.}}\label{FigBess}
\end{figure}

Notice the sums over the $\eta_l$ signs, inherited from the sum in \Ref{FT}. It was proved in \cite{grav3} that each $\eta_l$ configuration gives the same LO asymptotics in $1/j_0$.\footnote{The coefficients are different than when using the Gaussian state with the complex phase: the difference comes entirely from the phase, and it is due to the fact that the oscillating part of \Ref{psiB} is the real part of the oscillating part of \Ref{psiG}.  Because the state enters in both numerator and denominator of \Ref{W1}, the result with \Ref{psiB} coincides with the ratio of the real parts of the
numerator and denominator computed using \Ref{psiG}.} Therefore for the LO the sums over $\eta_l$ give an irrelevant redundancy which is cancelled by the denominator. The calculations of the LO were confirmed by the numerics, although the plots were not included in \cite{grav3}. Here we fill that gap and present the relevant plots in Figure \ref{FigBess}.
The left panel shows the numerical evaluation with only the coherent configuration with $\eta_l=1$ for all $l$, whereas the right panel shows the result after summing over all the possible $\eta_l$ signs. We see that the LO is the same, as claimed above, however the full non-perturbative evaluation is rather different. In particular, the coherent $\eta_l$ configuration is very similar to the Gaussian state calculation, whereas the sum over the $\eta_l$ creates interferences and introduces oscillations which make the global picture quite different. In particular, the oscillations also swamp the short scale peak.

To comment on these oscillations, let us look again at \Ref{FT}. From the Gaussian shape of its right hand side,
it is clear that this Fourier transform peaks the $\SU(2)$ class angle $\phi$ on both $\theta$ and $\pi-\theta$, with spread $\sqrt{\alpha/j_0}$.
Notice also that because of the restriction on the spins made above, now all $j_l$ are either integers, or integers shifted by a half, accordingly to $j_0$. Therefore taking $\theta$ to be either the internal or external dihedral angle leads to the same asymptotics, a very different situation from the complex phase of the Gaussian ansatz \Ref{psiG}, where only the right choice of angle in the boundary state leads to a good asymptotics in $1/j_0$.
This symmetry between the internal and external angle might look appealing at first, but it is in fact the origin of the messier non-perturbative behaviour of \Ref{WB} highlighted by the right panel of Figure \ref{FigBess}.

Let us recall that the kernel \Ref{10j}, being made of sines, is a product of terms each of which contains
two exponentials with opposite signs. This is a characteristic of the spin foam formalism, which appears also in the new models, and it is interpreted as the fact that the kernel sums over orientations of the spacetime manifold. When acting with a state like \Ref{psiG}, only one sign for each term contributes to the asymptotics, whereas the terms with the ``wrong'' signs are exponentially suppressed. This mechanism shows that a complex phase in the boundary is key to give an (observer-induced) orientation to the kernel. Our results show also that fixing this orientation is important for having a nice short scale picture without large oscillations.

\section{Conclusion and Outlook}\label{se:concl}

In this paper we numerically studied various aspects of the spin foam graviton propagator on a single 4-simplex, using the Barrett-Crane model for the dynamics. Although now replaced by the better-behaved EPR-FK-LS spin foam model, the Barrett-Crane model still provides an excellent model for testing many ideas, and furthermore it is already at the edge of present computational power.

Our results show that numerical techniques are very useful for studying the type of correlation described by the propagator. One can go beyond merely checking the analytic calculation of the leading order, and study higher order corrections and short scale physics. We showed how both are affected by the triangle amplitudes, corresponding to measure choices in the path integral. In particular, the trivial triangle amplitude minimizes the quantum corrections. The short scale picture presents the remarkable feature of a peak, followed by an inversed behaviour of the correlation. This result, already presented in \cite{letter}, is very suggestive that interesting physical effects can be studied by looking at these types of quantities. We showed how the peak depends on the parameter in the boundary state, and then compared the behaviour of this model with more realistic short scale scenarios arising from alternative approaches to non-perturbative quantum gravity.
Finally, we compared the usual Gaussian ansatz for the boundary state with the Bessel-based state.

As a final remark, our results also indicate that it is important to pour effort into overcoming the current limitations of numerical simulations, in particular to be able to treat non-factorized boundary states, as well as many 4-simplices.

\appendix
\section{Spin summation method}\label{app:summation}
The spin summation method introduced in Section~\ref{numericalsection}
and used in Section~\ref{nlosection} is a generalization of the
Christensen-Egan (CE) algorithm for efficient evaluation of the
$10j$-symbol~\cite{CEalg}. This generalization is described in full
detail%
\footnote{There is an important difference in notation
between~\cite{Khavkine} and this paper. The former uses twice-spins
(taking on only integer values) to
label irreducible representations of SU(2), while the latter uses spins
(taking on also half-integral values).}, %
in~\cite{Khavkine}. Section~3.3
of~\cite{Khavkine} is most relevant to this paper, as it treats the case
of a single Barrett-Crane vertex contracted with a factored boundary
state. Below we give a brief overview of these techniques, emphasizing
their innovations compared to previously known ones.

To start, we summarize the CE algorithm. As discussed in the original
paper, the $10j$-symbol can be written as a contraction of $\SU(2)$
Clebsch-Gordan (CG) coefficients. Each CG coefficient is a finite-dimensional
rank 3 $\SU(2)$ tensor. As such, one can naively try to
evaluate the $10j$-symbol by multiplying all the CG tensor
components and then summing over each contracted index. Unfortunately,
the sums obtained in this way are prohibitively large. The CE algorithm
makes two significant advances to make the evaluation of these sums
feasible.

First, it uses $\SU(2)$ recoupling identities to eliminate the
$\SU(2)$-representation index sums in exchange for a smaller number of
intermediate irrep label sums (spin sums). At this stage, the CG
tensors have been eliminated in favor of $\SU(2)$ $\Theta$- and
$6j$-symbols (see~\cite{CEalg} for details), functions of $3$ and $6$
spins respectively. These, in turn, can be formally seen as infinite-dimensional%
	\footnote{Even though these tensors are infinite-dimensional,
		sufficiently many of their components vanish to make the relevant
    sums finite. } %
tensors of rank 3 and 6, even though their indices (spins) do not
necessarily label basis vectors in any linear space. To reiterate
briefly, $\SU(2)$ recoupling identities are used to express the
contraction of many CG tensors, with a large number of dummy indices, as
a contraction of some different tensors, now with a smaller number of
dummy indices.

Notice that some tensor contractions can be evaluated more efficiently
than others. In particular, the trace of a product of $k$ $n\times n$
matrices is a contraction of $k$ $n$-dimensional, rank 2 tensors,
involving $k$ dummy indices. A naive estimate for the number of
operations needed to evaluate this contraction as a sum over each dummy
index gives $O(n^k)$. On the other hand, standard matrix multiplication
algorithms can evaluate the same product trace in $O(k n^3)$ steps, a
significant gain in efficiency for $k>3$. The second significant advance
of the CE algorithm is to recognize that part of the expression for the
$10j$-symbol as a contraction of $\theta$- and $6j$-symbols has
precisely the form of the trace of a product of matrices, which can then
be evaluated efficiently.

Inserting the CE spin sum expression for the Barrett-Crane
$10j$-symbol into~\Ref{W} gives a spin sum expression for the area
correlations $W$. Thinking of summed spins as contracted tensor
indices, this expression can be thought of as another large tensor
contraction. At first glance, its evaluation would require a
prohibitively large number of operations, however it is amenable to
the same kind of analysis as in the CE algorithm.

Again, two steps are essential in making the spin summation method
feasible. First, one notices that this new tensor contraction
still has the matrix product trace structure of the original CE
algorithm, however only if the boundary state is factorable, as
in~\Ref{psiFact}. This condition is where the technical restriction to
factorable boundary states appears. Second, the normally dense matrices
in this product trace can be decomposed into sparse, structured factors.
This factorization is achieved using $\SU(2)$ recoupling identities
involving products of $6j$-symbols.
The reason this factorization speeds up the calculations is that
multiplication of sparse, structured matrices (diagonal, banded and
other types) is significantly faster than that of dense ones.


\end{document}